\documentclass[reprint,superscriptaddress,floatfix]{revtex4-1}
\usepackage[utf8]{inputenc}
\usepackage{graphicx}
\usepackage{physics}
\usepackage{hyperref}

\usepackage{float}
\usepackage{xcolor}
\usepackage{textcomp}
\usepackage{xfrac}
\usepackage{nicefrac}

\begin{document}

\title{Machine learning active-nematic hydrodynamics}

\author{Jonathan Colen}
\thanks{These authors contributed equally}
\affiliation{Department of Physics, University of Chicago, Chicago, Illinois, 60637, U.S.A.}
\affiliation{James Franck Institute, University of Chicago, Chicago, Illinois, 60637, U.S.A.}
\author{Ming Han}
\thanks{These authors contributed equally}
\affiliation{James Franck Institute, University of Chicago, Chicago, Illinois, 60637, U.S.A.}
\affiliation{Pritzer School of Molecular Engineering, University of Chicago, Chicago, Illinois, 60637, U.S.A}
\author{Rui Zhang}
\affiliation{Pritzer School of Molecular Engineering, University of Chicago, Chicago, Illinois, 60637, U.S.A}
\author{Steven A. Redford}
\affiliation{James Franck Institute, University of Chicago, Chicago, Illinois, 60637, U.S.A.}
\affiliation{Graduate Program in Biophysical Sciences, University of Chicago, Chicago, Illinois, 60637, U.S.A.}
\author{Linnea M. Lemma}
\affiliation{Department of Physics, Brandeis University, Waltham, Massachusetts, 02454, U.S.A.}
\affiliation{Department of Physics, University of California at Santa Barbara, Santa Barbara, California, 92111, U.S.A.}
\author{Link Morgan}
\affiliation{Department of Physics, University of California at Santa Barbara, Santa Barbara, California, 92111, U.S.A.}
\author{Paul V. Ruijgrok}
\affiliation{Department of Bioengineering, Stanford University, Stanford, California, 94305, U.S.A.}
\author{Raymond Adkins}
\affiliation{Department of Physics, University of California at Santa Barbara, Santa Barbara, California, 92111, U.S.A.}
\author{Zev Bryant}
\affiliation{Department of Bioengineering, Stanford University, Stanford, California, 94305, U.S.A.}
\affiliation{Department of Structural Biology, Stanford University Medical Center, Stanford, California, 94305, U.S.A.}
\author{Zvonimir Dogic}
\affiliation{Department of Physics, University of California at Santa Barbara, Santa Barbara, California, 92111, U.S.A.}
\author{Margaret L. Gardel}
\affiliation{Department of Physics, University of Chicago, Chicago, Illinois, 60637, U.S.A.}
\affiliation{James Franck Institute, University of Chicago, Chicago, Illinois, 60637, U.S.A.}
\author{Juan J. De Pablo}
\thanks{vitelli@uchicago.edu; depablo@uchicago.edu}
\affiliation{Pritzer School of Molecular Engineering, University of Chicago, Chicago, Illinois, 60637, U.S.A}
\affiliation{Center for Molecular Engineering, Argonne National Laboratory, Lemont, Illinois, 60439, USA}
\author{Vincenzo Vitelli}
\thanks{vitelli@uchicago.edu; depablo@uchicago.edu}
\affiliation{Department of Physics, University of Chicago, Chicago, Illinois, 60637, U.S.A.}
\affiliation{James Franck Institute, University of Chicago, Chicago, Illinois, 60637, U.S.A.}

\begin{abstract}
    Hydrodynamic theories effectively describe many-body systems out of equilibrium in terms of a few macroscopic parameters. However, such hydrodynamic parameters are difficult to derive from microscopics~\cite{Marchetti2013}. Seldom is this challenge more apparent than in active matter where the energy cascade mechanisms responsible for autonomous large-scale dynamics are poorly understood~\cite{Giomi2015,AditiSimha2002,doostmohammadi2017onset,Alert2020,wensink2012meso}. Here, we use active nematics~\cite{sanchez2012spontaneous,Keber2014a,ellis2018curvature,Kumar2018,Duclos2020} to demonstrate that neural networks~\cite{Lecun2015,Schmidhuber2015} can extract the spatio-temporal variation of hydrodynamic parameters directly from experiments. Our algorithms analyze microtubule-kinesin~\cite{sanchez2012spontaneous,decamp2015orientational,Lemma2019a,Hardouin2019} and actin-myosin~\cite{Kumar2018,zhang2019structuring} experiments as computer vision problems. Unlike existing methods, neural networks can determine how multiple parameters such as activity and elastic constants vary with ATP and motor concentration. In addition, we can forecast the evolution of these chaotic many-body systems solely from image-sequences of their past by combining autoencoder and recurrent networks with residual architecture~\cite{he2015resnet}. Our study paves the way for artificial-intelligence characterization and control of coupled chaotic fields in diverse physical~\cite{wu2017transition,zhang2019structuring} and biological~\cite{saw2017topological,duclos2017topological,maroudas2020topological,mueller2019emergence} systems even when no knowledge of the underlying dynamics exists.
\end{abstract}

\maketitle

Machine learning holds great promise as a tool capable of revolutionizing traditional approaches to mathematical and computational modelling in the physical sciences~\cite{mehta2019highbias,Carleo2019}. Notable successes include learning how to identify phases of matter~\cite{Carrasquilla2017} or
recognize glasses \cite{schoenholz2016structural,bapst2020unveiling}, and revisiting fundamental concepts like the renormalization group ~\cite{mehta2014exact,Koch-Janusz2018}. However, the \textit{experimental} validation of machine learning as a tool capable of addressing otherwise intractable challenges is still in its infancy~\cite{Cubuk2017structure,radovic2018machine}. Machine learning techniques can also transform the way we look at and model
many-body dynamical systems~\cite{Marchetti2013}. Active nematics~\cite{Ramaswamy2010,Doostmohammadi2018} are a paradigmatic example of this class of systems.
Their chaotic dynamics are sufficiently well characterized to be a reliable benchmark 
and yet hard to predict owing to their far from equilibrium nature. 

\begin{figure}[h]
    \centering
    \includegraphics[width=0.465\textwidth]{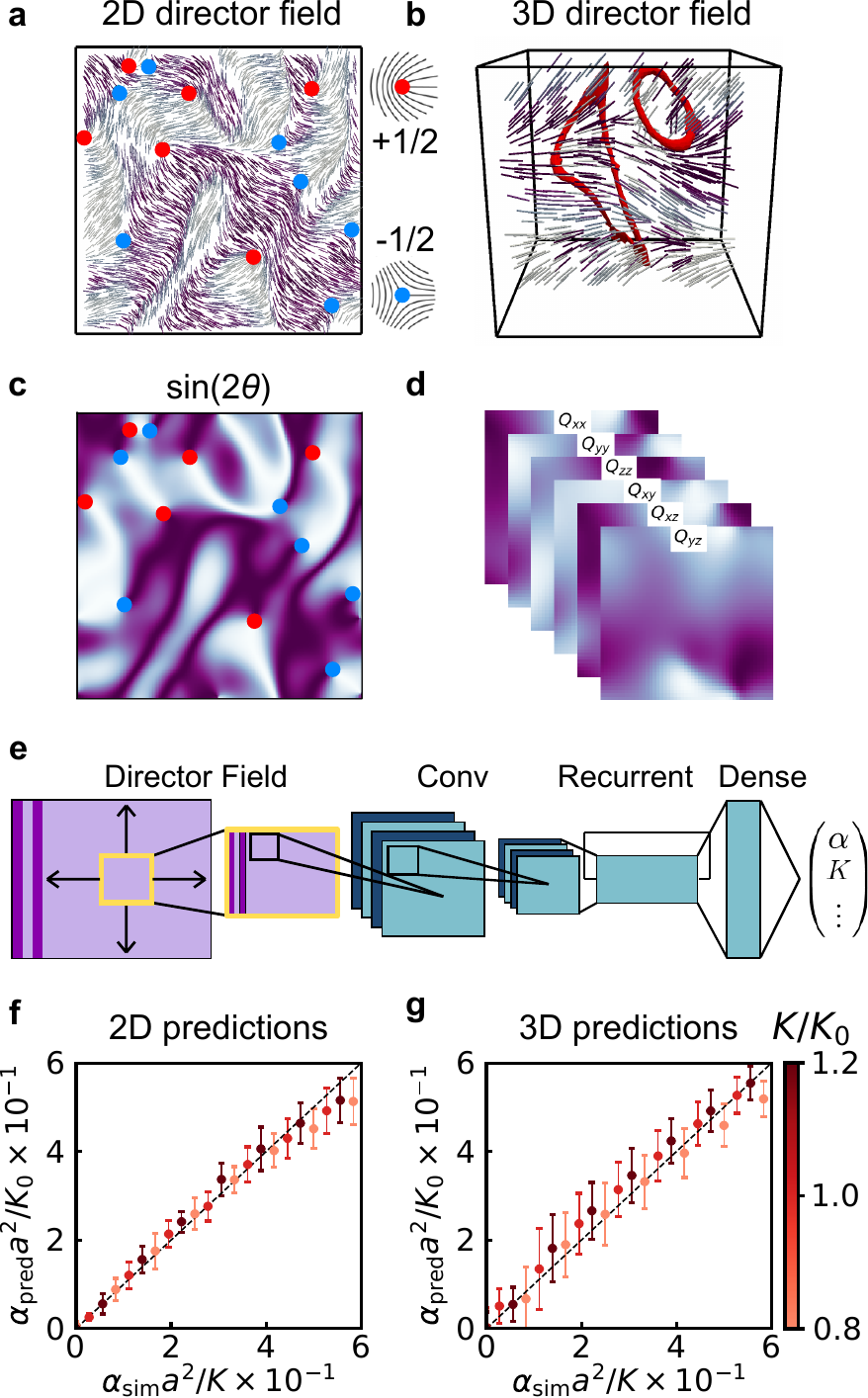}
    \caption{\textbf{Machine learned hydrodynamic parameters in Lattice-Boltzmann simulations.} \textbf{a, b.} Raw nematic director fields in two (\textbf{a}) and three (\textbf{b}) dimensions. $\sfrac{+1}{2}$ and $\sfrac{-1}{2}$  defects in 2D are marked as red and blue dots, respectively. Disclination loops are indicated in red. \textbf{c, d.} Smooth representations of the director field used by the network. In two dimensions, the network uses $\sin 2\theta$ where $\theta$ is the angle of the director field. In 3D, the network uses the tensor $Q_{ij} = n_i n_j - 1/3$. Color indicates the magnitude of these smooth representations. \textbf{e.} Schematic of neural network architecture. The full input images are divided into patches, which are then fed into a set of convolutional filters, a LSTM recurrent layer, and a fully connected dense layer. The outputs are ensemble averaged into a final estimate for hydrodynamic parameters. \textbf{f, g.} Predictive accuracy of rescaled dimensionless activity in simulation data in 2D and 3D at different values of $K$. Networks were trained at $K=K_0$. Units such as $K_0$ are listed in Methods.}
    \label{fig:parameter_estimation}
\end{figure}

Nematic liquid crystals are anisotropic materials composed, for example, of microscopic filaments whose average orientation at position $\textbf{r}$ is measured by the director field $\textbf{n}(\textbf{r})$. Representative images of the director field are shown in Fig.~\ref{fig:parameter_estimation}a-b.
In equilibrium nematics, the filaments tend to align along the director. Gradients of $\textbf{n}(\textbf{r})$ are penalized by the Frank free-energy density
\begin{equation}
   f_d = \frac{K}{2} (\partial_i n_j)^2
    \label{eq:free_energy}
\end{equation}
where all the elastic constants are set equal to $K$, for simplicity (see Methods). 
The study of nemato-hydrodynamics requires keeping track of both the filaments' local velocity field $\textbf{v}(\textbf{r})$ and orientation via the director field $\textbf{n}(\textbf{r})$ \cite{Marchetti2013,Doostmohammadi2018}.
These fields are coupled by non-linear evolution equations whose complexity poses a stringent test for our machine learning methods (see Methods).

The introduction of microscopic energy sources into a system with orientational order can create non-equilibrium dynamical states called active nematics.~\cite{Marchetti2013,Ramaswamy2010,Doostmohammadi2018} A common example is cytoskeleton filaments with molecular motors that promote inter-filament sliding. In  such  media,
the microscopically injected energy cascades across length scales eventually leading to the proliferation of topological defects ~\cite{sanchez2012spontaneous,Giomi2013,AditiSimha2002,Voituriez2005,shankar2018defect,shankar2019hydrodynamics}. 
The ensuing chaotic dynamics that occurs above a well defined activity threshold is sometimes referred to as active nematic turbulence~\cite{AditiSimha2002,Giomi2015,Doostmohammadi2018,Alert2020,wensink2012meso,hatwalne2004rheology}. Note, however, that the underlying cascade mechanism in active nematics is different from ordinary turbulence and still poorly understood. 

The chaotic dynamics is driven by the active stress 
\begin{equation}
    \mathbf{\sigma}^{a}_{ij} = \alpha\,n_i n_j
    \label{eq:active_stress}
\end{equation}
where $\alpha$ is an \textit{a priori} unknown macroscopic activity coefficient ~\cite{AditiSimha2002,Marchetti2013}. How hydrodynamic parameters, such as $\alpha$ and $K$, \textit{independently} vary with microscopic energy transduction, e.g., ATP or motor concentrations, remains an open question. Direct measurements that probe these parameters independently are difficult to devise owing to the non-quiescent nature of the chaotic steady state. Furthermore, most responses depend on the competition between active stresses, that promote director or velocity gradients, and viscoelastic stresses that resist them. As a consequence of this interplay, experimental measurements often access only non-trivial \textit{combinations} of hydrodynamic parameters~\cite{Kumar2018,Lemma2019a}. Furthermore, the task of deriving these parameters from microscopic models is daunting, prompting us to seek approaches that bypass standard coarse-graining.  

To make progress, we recast the task of estimating multiple hydrodynamic parameters as a computer-vision problem that can be effectively addressed by artificial intelligence.
We start by generating a library of director fields (Fig.~\ref{fig:parameter_estimation}a-b) for a wide range of activity in two and three dimensions using
Lattice-Boltzmann  simulations~\cite{Marenduzzo2007,Kumar2018} (see Methods).
Using this library, we train neural networks on smooth representations of $\mathbf{n}$, see Fig.~\ref{fig:parameter_estimation}c-d. The neural network architecture, shown schematically in Fig.~\ref{fig:parameter_estimation}e, contains (i) a single convolutional layer used for image processing, (ii) a recurrent layer that captures the system dynamics, and (iii) a dense layer that identifies the hydrodynamic parameters.  

We first apply this scheme to estimate a single parameter: the rescaled dimensionless activity $\alpha / K \times a^2$, 
where $a$ denotes the pixel or voxel size for the director field image. 
Comparison of the machine learning predictions for the activity with the known values of $\alpha$ (the \textit{ground truth}, in machine learning parlance)
reveals good agreement for both two-dimensional (2D) and three-dimensional (3D) active nematics (Fig.~\ref{fig:parameter_estimation}f-g). 
Although these networks are trained on data generated at a single value of $K = K_0$, their predictions of the rescaled activity parameter are remarkably accurate even for samples where $K$ differs from $K_0$.

For 2D active nematics, hydrodynamic theories suggest that $\alpha$ can be estimated from the characteristic length $\ell_{\text{d}} \propto \sqrt{K/\alpha}$
obtained from dimensional analysis by balancing the right hand sides of Eqs.~\eqref{eq:free_energy} and \eqref{eq:active_stress}, 
which have the same units.
In 2D samples, $\ell_{\text{d}}$ can be interpreted physically as the average spacing between point disclinations, topological defects with index $\sfrac{+1}{2}$ and $\sfrac{-1}{2}$ shown in Fig.~\ref{fig:parameter_estimation}a as red and blue dots, respectively~\cite{Marchetti2013,Giomi2013}. However, this heuristic approach does not work if disclinations are not present in the field of view, a common occurrence at low activity. In addition, this approach does not generalize to 3D nematics whose dominant excitations are charge-neutral disclination loops ~\cite{Duclos2020} (Fig.~\ref{fig:parameter_estimation}b). 
It is unclear how the structure and dynamics of these loops vary with activity. Both of these limitations are readily overcome by our machine learning approach (Figs.~\ref{fig:parameter_estimation}f-g). Instead of extracting topological defects, which would require multiple convolutional layers, our networks (which contain only a single convolutional layer) simply exploit local spatial fluctuations of the director field.

\begin{figure*}
    \centering
    \includegraphics[width=0.9\textwidth]{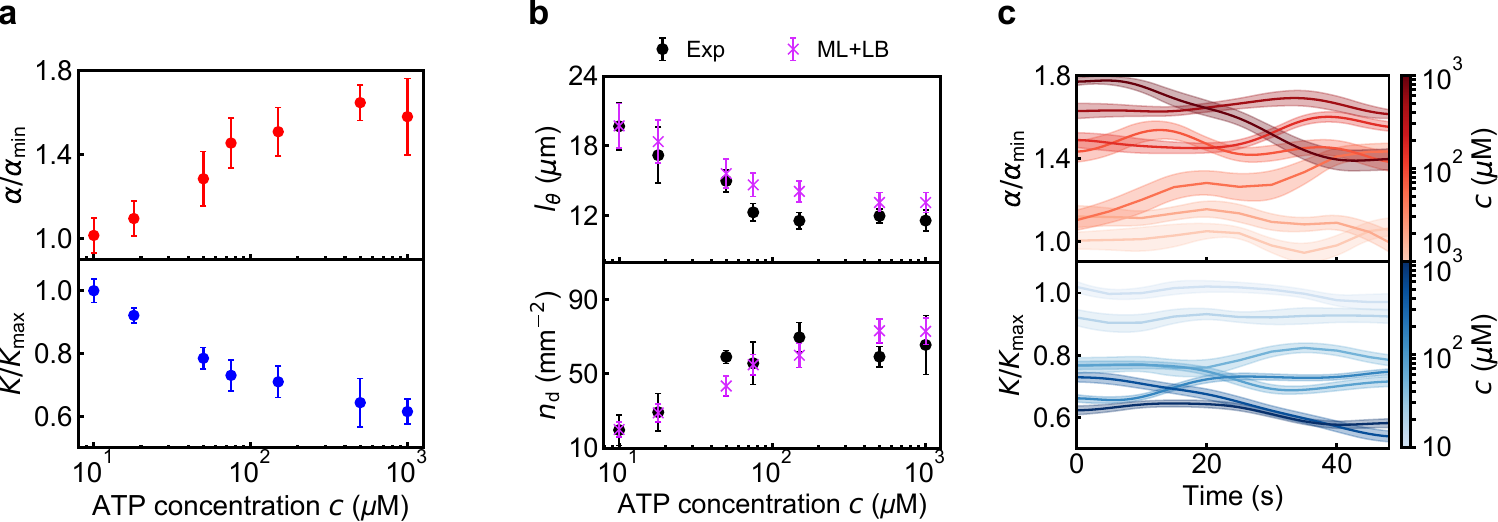}
    \caption{\textbf{Multi-parameter estimation in microtubule-kinesin experiments}. \textbf{a.}
     Dependence of spatio-temporally averaged activity and elastic modulus on ATP concentration. 
    Here, $\alpha_{\text{min}}$, $K_{\text{max}}$  are the time-averaged predicted activity and elastic modulus at the lowest level of ATP concentration $c_\text{min} = 10$ $\mathrm{\mu}$M.
    \textbf{b.} Comparison of director field correlation length $l_{\theta}$ and defect spacing $n_{\text{d}}$ in experiments (Exp) and machine learning informed Lattice-Boltzmann simulations (ML+LB).
    \textbf{c.} Simultaneous prediction of activity and elastic modulus over time at different levels of ATP concentration. The shaded regions represent the standard-error of spatio-temporal fluctuations in the machine learning predictions. ATP concentration $c$ is indicated by the color bar.}
    \label{fig:experimental_data}
\end{figure*}

While neural networks designed for single-parameter estimation can already address challenges intractable by existing methods, they still predict only combinations of parameters such as $\alpha / K$, very much like the disclination spacing $\ell_\text{d}$ in 2D samples with sufficiently high activity. To extract $\alpha$ independently one would still need to resort to \textit{ad hoc} assumptions like the independence between $K$ and $\alpha$ which is not always experimentally valid. By contrast, our neural networks can perform multi-parameter estimation without the need to devise a set of challenging experiments that disentangle the parameters' interdependence. Supplementary Figure~\ref{fig:multi2d} illustrates how our machine learning models can predict $\alpha$ and $K$ independently in Lattice-Boltzmann simulation tests. 

So far we have demonstrated the power of these machine learning algorithms on numerical data. We now apply our multi-parameter estimation to experiments performed on microtubule-kinesin systems~\cite{sanchez2012spontaneous} (see Methods). 
It is known that the rescaled activity $\alpha / K$ increases with ATP concentration~\cite{Lemma2019a} which controls the stepping speed of kinesin motors~\cite{Schnitzer1997}. Here, we proceed to disentangle the dependence of $\alpha$ and $K$ on the ATP concentration, $c$.
Inspection of Fig.~\ref{fig:experimental_data}a reveals that the spatio-temporally averaged activity, $\alpha$, predicted by our machine learning algorithms increases with $c$ while the elastic modulus, $K$, decreases. Similar results for 3D microtubule-kinesin nematics and 2D actin-myosin systems are shown in Supplementary Figs.~\ref{fig:3d_predictions}--\ref{fig:actin-multi2d}. 

Our machine learning approach to multi-parameter estimation differs from the traditional method of ``curve-fitting" experimental data. Both start with a theory whose parameters are to be determined from data. However, the traditional method also requires that one can (i) identify special conditions where the underlying theory is solvable and (ii) parameterize the solution for a measurable observable in terms of the sought-after hydrodynamic coefficients. 
In contrast, our neural networks can be trained on data obtained using whatever conditions are experimentally available. We stress that neural networks do not simply employ lookup tables. For a physical field like $\textbf{n}(\textbf{r}, t)$, constructing lookup tables is impractical as the dimension of data (number of pixels or voxels here) often exceeds the number of data points. Instead, our neural networks \textit{learn} the optimal high-dimensional manifold that encodes the connection between all possible realizations of the field, e.g. $\textbf{n}(\textbf{r}, t)$, and the corresponding hydrodynamic parameters. 

Although the active nemato-hydrodynamic theory adopted in our Lattice-Boltzmann simulations is widely accepted, its validity has not been thoroughly tested against experiments. 
This requires precisely the simultaneous identification of absolute values of hydrodynamic parameters that has so far been missing. Here, we use the machine learning predicted parameters at different ATP concentrations to run several rounds of Lattice-Boltzmann simulations and compare their results to experiments. 
As the chaotic nature of active nematics makes exact director field comparisons between machine learning  informed Lattice-Boltzmann simulations and experiments unreliable over long times, we instead compute properties of the dynamical steady state.
Using the spatial correlation function $C_s(\textbf{r})$ for the director field (see Methods), we define the correlation length $\ell_{\theta}$ such that $C_s(\ell_{\theta}) = 1/2$. 
We find that both the average correlation length and the defect density $n_{\text{d}}$ calculated from machine learning informed Lattice-Boltzmann simulations match the ground truth probed directly in experiments for a wide range of ATP concentrations (Fig.~\ref{fig:experimental_data}b), hence validating the hydrodynamic theory.

Until now, the experimental characterization of time-averaged properties has not departed from the standard continuum theory under the assumption that the hydrodynamic parameters are constant.
However, our machine learning methods can provide a rare glimpse into the hitherto neglected time fluctuations of $\alpha$ and $K$ that are not captured by the hydrodynamics theory with constant coefficients. Figure.~\ref{fig:experimental_data}c shows an example of time series for $\alpha$ and $K$. The mean values are plotted as solid lines and their uncertainties are marked with shaded regions, over a wide range of ATP concentrations (denoted by color bars).
The variations observed in experiments, especially at the highest ATP concentrations, are markedly more pronounced than those observed in simulations (Supplementary Fig.~\ref{fig:fluctuations_const}) where they arise solely from uncertainties in the machine learning predictions themselves. 
This machine-learned evidence suggests that a non-linear fluctuating hydrodynamic theory may better explain our experimental observations. Heuristically, the strong disruption of fiber alignment at large activity can trigger motor detachment-reattachment events causing the time modulation of $\alpha$ inferred by our algorithms. This microscopic scenario needs to be carefully tested by detailed biochemical experiments that lie beyond the machine learning methodology exposed here.

\begin{figure*}
    \centering
    \includegraphics[width=0.88\textwidth]{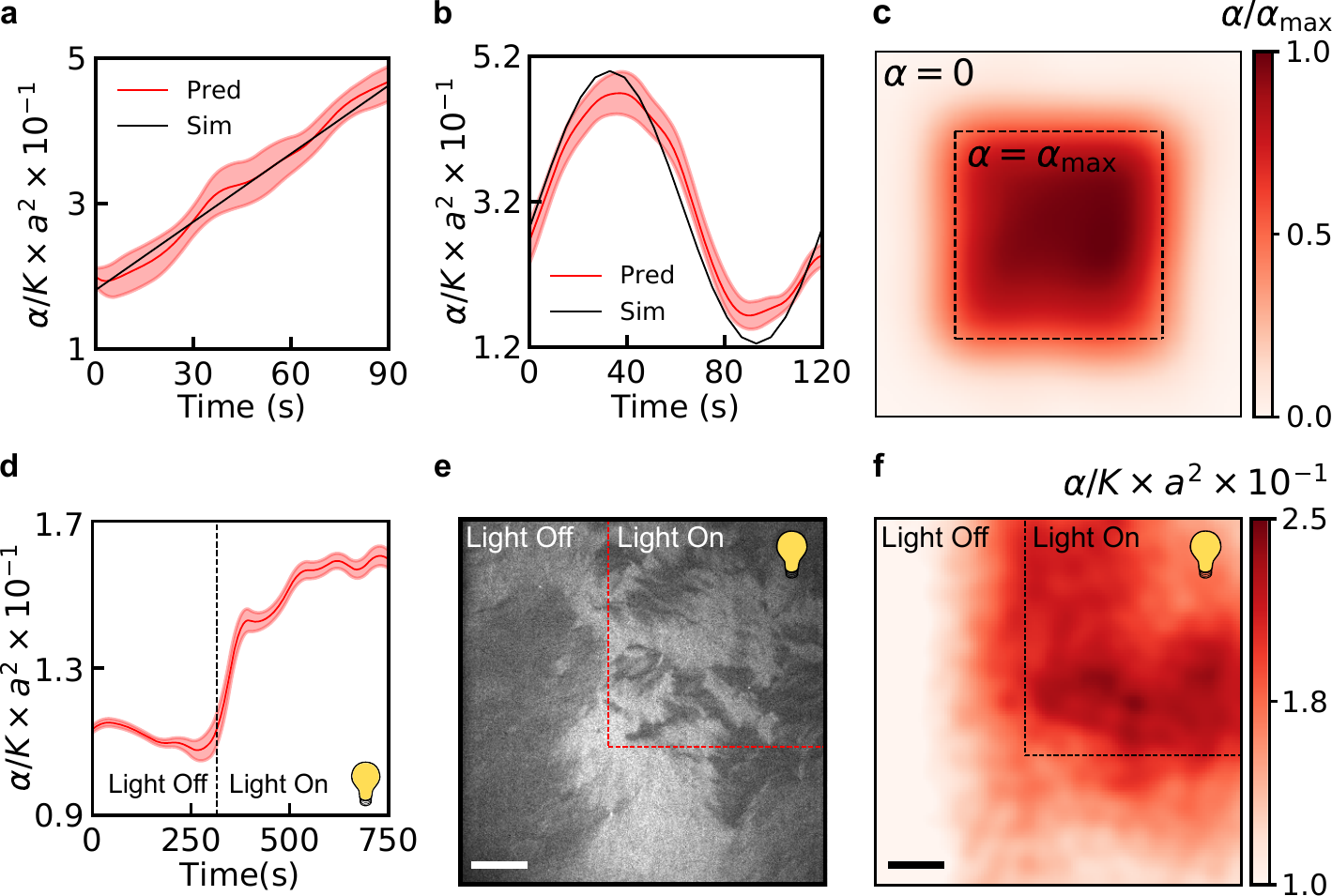}
    \caption{\textbf{Machine-learned spatio-temporal activity in Lattice-Boltzmann simulations and actin-myosin experiments.} \textbf{a-b.} Machine learning predicted activity on simulations with prescribed time-varying activity coefficients, linear (\textbf{a}) and sinusoidal (\textbf{b}). \textbf{c.} Machine learning predicted activity on simulations where the central square (dashed line) is activated.
    \textbf{d.} Machine learning predicted activity vs. time on actin-myosin experiments where myosin motors are controlled through light-activated gearshifting. The dashed line indicates when light is switched on \textbf{e-f.} Direct image (\textbf{e}) and machine learning predicted spatial activity profile (\textbf{f}) of a selectively illuminated actin nematic with light-activated gearshifting motors. 
    For \textbf{e-f.} the experimental data is the dataset reported in Fig.~1 of Zhang \textit{et al.} \cite{zhang2019structuring}. Data for \textbf{d} are from the current study, following the approach used in \cite{zhang2019structuring}. Scale bars, 20 $\mu$m. }
    \label{fig:time_dependence}
\end{figure*}

We now apply our machine learning models to situations in which activity is engineered to \textit{deliberately} vary in both time and space. We test this activity control scenario first in Lattice-Boltzmann simulations, where we prescribe spatio-temporal patterns of $\alpha(\textbf{r},t)$~\cite{zhang2019structuring}.
Remarkably, neural networks trained on the data with constant activity can still accurately estimate a time-varying activity coefficient as shown in Fig.~\ref{fig:time_dependence}a-b where linear and sinusoidal activity profiles are probed.  
Since small director-field patches are sufficient to generate reliable predictions, we can generate a spatial activity map of $\alpha(\textbf{r}, t)$ by applying our neural networks locally to each patch composing an image.
By doing this, we are able to discern prescribed spatial activity patterns in Lattice-Boltzmann simulations, as demonstrated in Fig.~\ref{fig:time_dependence}c where activity is non-zero only in the central square.

In experiments on actin-myosin nematics~\cite{Kumar2018}, it is possible to alter the speed of some specialized molecular motors via selective exposure to light~\cite{Nakamura2014RemoteGearshifting}.
This phenomenon, informally called gear-shifting (see Methods), allows for precise $spatio$--$temporal$ control of active stresses~\cite{zhang2019structuring}.
Inspection of Fig.~\ref{fig:time_dependence}d and Supplementary Movie~1 reveals that our machine learning models can successfully identify the marked increase in activity that occurs as light is turned on (indicated by the dashed line in Fig.~\ref{fig:time_dependence}d).
Furthermore, our approach can identify the activity changes that occur in selectively illuminated \textit{spatial} domains in these systems, see Fig.~\ref{fig:time_dependence}e-f and Supplementary Movie~1. 
The success of our machine learning model in identifying spatio-temporally varying activity demonstrates its potential for the control of engineered active materials and the inference of biochemical processes that take place at the microscopic level (Supplementary Table ~I).

\begin{figure*}
    \centering
    \includegraphics[width=0.9\textwidth]{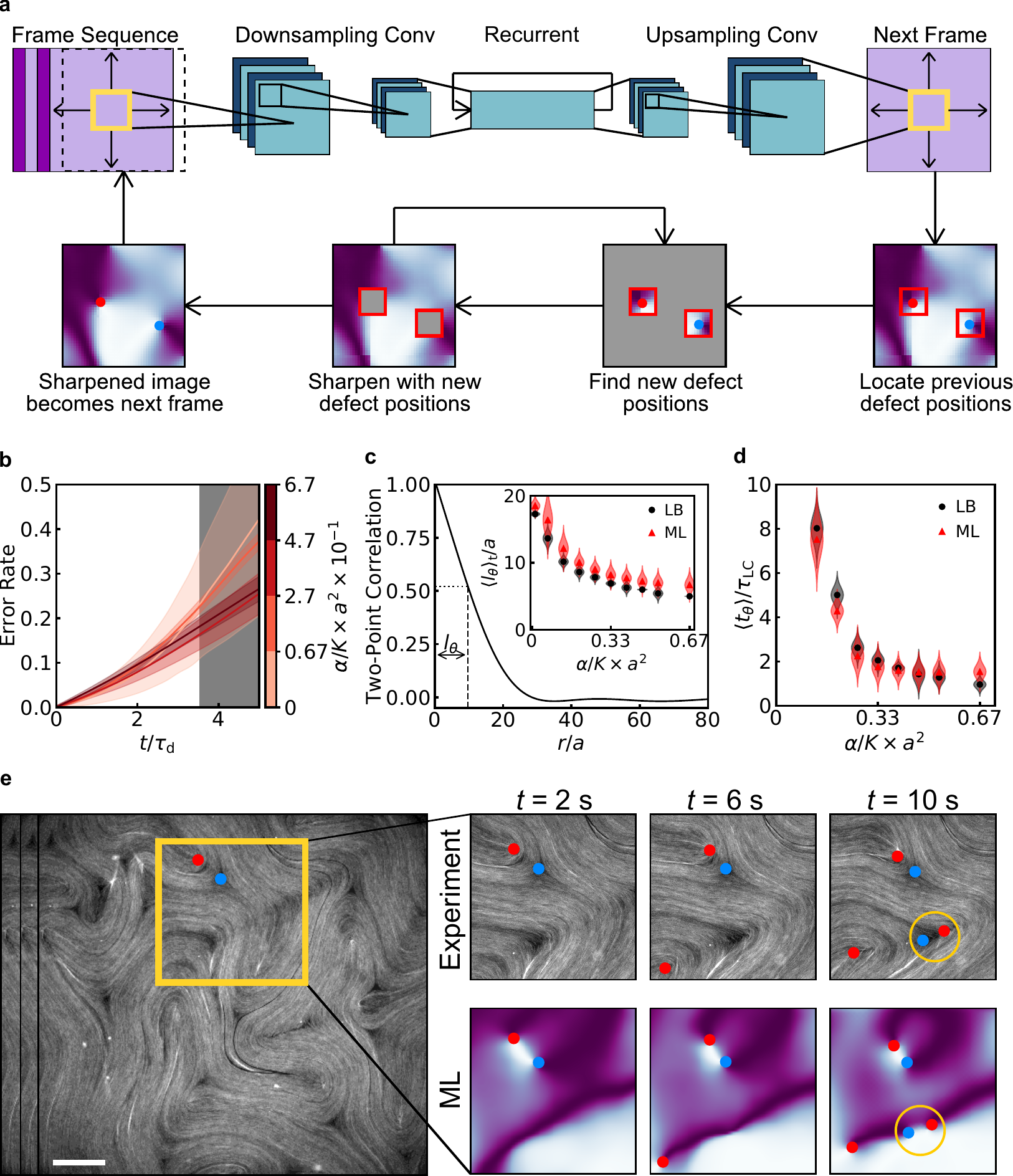}
    \caption{\textbf{Neural-networks as surrogate models of time evolution}. \textbf{a.} Schematic for predicting the time evolution of an active nematic. A sequence of images is fed into a modified convolutional autoencoder architecture (with a recurrent layer inserted in the middle). The output is sharpened by first using relaxational dynamics to update defect positions, and then using the updated defect positions to sharpen the entire director field. This procedure is iterated and the sharpened image is used as the next frame.
    \textbf{b.} Pixel-wise error rate $1 - \left<| \textbf{n}_{\text{ML}} \cdot \textbf{n}_{\text{LB}} |\right>$ of the predictive model versus time, for different groups of activity in Lattice-Boltzmann simulations. The gray area shows regions beyond the Lyapunov time for the Lattice-Boltzmann simulations (see Methods). Here $\tau_{\text{d}} = \eta / \alpha$, the characteristic defect lifetime. \textbf{c.} Comparison of time-averaged correlation length in machine learning and Lattice-Boltzmann simulations. 
    \textbf{d.} Comparison of average director field correlation time $t_{\theta}$ in machine learning and Lattice-Boltzmann simulations.
    \textbf{e.} A defect nucleation event as seen in experiment and as predicted by the machine learning model trained on experimental data. machine learning predictions depict the magnitude of $\sin (2\theta)$, where $\theta$ is the angle of the director field. $\sfrac{+1}{2}$ and $\sfrac{-1}{2}$  defects are marked as red and blue dots, respectively. Scale bar 100 $\mu$m. }
    \label{fig:time_evolution}
\end{figure*}

We now ask: can neural networks learn the evolution of chaotic many-body systems solely from image-sequences of their past? The canonical approach to quantitative modelling in the physical sciences relies on writing down equations and then solving them, analytically or via simulations, to make predictions. 
In what follows, we use the word \textit{machine-learning model} to denote something very different.
Instead of solving the equations, 
we train neural networks on data and then ask them to forecast the future behavior of the chaotic system. 
An advantage of this approach is that these algorithms can be trained directly on experimental data. 
If reliable simulations are available, they can be used for training and testing, but this is by no means required.

To implement this forecasting approach in the context of active nematics, we iterate the following two steps.
First, we perform next-frame predictions using a neural network that does not know anything about the physics of the system. 
Second, we reduce any noise generated in the previous step by applying to each frame a physically-motivated sharpening algorithm, which minimizes the elastic energy but does not know anything about the active forces driving the non-equilibrium dynamics (see Methods). 

Our time-evolution neural network is a modification of the autoencoder architecture, a popular tool in the computer vision community. 
A traditional autoencoder learns to compress an image to a feature vector which is then used to reconstruct the image.
In our network, we insert a recurrent layer in between the encoder and the decoder to learn the system dynamics.
Crucial for insuring high performance is the following algorithmic trick: a residual architecture~\cite{he2015resnet}  is used in the recurrent layer to capture the \textit{difference} between frames, rather than the images themselves. More broadly, we expect the use of residual architecture to be helpful whenever our approach is applied to other systems whose underlying dynamical processes are governed by differential equations.
As illustrated in Fig.~\ref{fig:time_evolution}a, 
the network encodes a time series of director-field images into a sequence of feature vectors. Next, it uses them to predict the future state of the system, and finally decodes this state back into a director-field image (Fig.~\ref{fig:time_evolution}a). For large systems, the director field is divided into small overlapping domains. Machine learning predictions are made within each domain and then stitched into a final prediction of the next director field configuration.

Given a particular nematic configuration, our algorithm can reliably learn the spatio-temporal evolution of the director field including singular events such as defect annihilation and nucleation (see Supplementary Fig.~\ref{fig:defect_nucleation} and corresponding Supplementary Movie~2).
To systematically evaluate the accuracy of our machine learning predictions, we first compare the time-evolved director fields generated by machine learning and Lattice-Boltzmann simulations pixel by pixel. 
Such pixel-wise comparison is only meaningful within the Lyapunov time, the characteristic timescale after which a non-linear dynamical system becomes chaotic. Inspection of Fig.~\ref{fig:time_evolution}b shows that the pixel-wise error rate of the predicted director field $1 - \left<|\textbf{n}_{ML} \cdot \textbf{n}_{LB}|\right>$ remains small within the Lyapunov time, which we find is equal to $t_\lambda \sim 3.6 \tau_\text{d}$ ($\tau_\text{d}$ denotes the average defect lifetime, see Methods).
Beyond the Lyapunov time (shaded region in Fig.~\ref{fig:time_evolution}b), even the Lattice-Boltzmann simulations are unreliable at the pixel-wise level due to numerical precision.

In order to evaluate the predictive accuracy of our machine learning methods for even longer times, we turn to properties of the dynamical steady state such as the director field correlation length $\ell_{\theta}$ defined before and the correlation time $t_{\theta}$ defined by setting the time correlation function $C_{t} (t_{\theta}) = 1/2$ (see Methods). 
Previous numerical studies have shown that the quantity $\ell_{\theta}$ is proportional to $\sqrt{K / \alpha}$~\cite{Hemingway2016}. 
When comparing the predictions of our machine learning model against Lattice-Boltzmann simulations, we find that machine learning correctly captures the activity dependence of the characteristic length $\ell_{\theta}$, see Fig.~\ref{fig:time_evolution}c. 
We stress that while $\ell_\theta$ at steady state is plotted in both Fig.~\ref{fig:experimental_data}b and Fig.~\ref{fig:time_evolution}c, the former is generated from Lattice-Boltzmann simulations with machine-learned parameters whereas the latter is generated solely using our time-evolution neural network.
Furthermore, our networks also reproduce the same activity dependence for $t_{\theta}$ as the Lattice-Boltzmann simulations (Fig.~\ref{fig:time_evolution}d), suggesting that they have learned to reproduce the correct dynamics expected at each level of activity. 

We now apply our machine learning time-evolution algorithm to experiments on microtubule-kinesin nematics. 
After training a neural network \textit{exclusively} on experimental data, we successfully forecast the time evolution of the nematic director over time scales that encompass various defect nucleation/annihilation events, see Fig.~\ref{fig:time_evolution}e for an example. 
In Supplementary Movie~3 we show the corresponding experimental video next to the machine learning generated one. 
Inspection of these movies show no discernible differences in the defect dynamics between experiments and machine learning predictions. Similar agreement is obtained when our time-evolution neural networks are trained on Lattice-Boltzmann simulation data (Supplementary Movie~3). 
To evaluate the long-term validity of our predictions
beyond specific realizations, we systematically check (as we did in Fig.~\ref{fig:experimental_data}b and Fig.~\ref{fig:time_evolution}c) that the steady-state values of $\ell_{\theta}$ and $n_{\text{d}}$ extracted from the full machine learning predicted nematic director are in good agreement with the experimental ground truth over a wide range of ATP concentrations (Supplementary Fig.~\ref{fig:mt_length_scales}). 

When trained directly on experimental data, our neural networks can forecast the future without theoretical knowledge of the underlying dynamics. This scenario is particularly intriguing for experimental systems that, unlike active nematics, lack a quantitative descriptions of the coarse grained dynamics. In addition, the simplicity of our time-evolution neural networks make it a suitable tool to implement artificial-intelligence informed control of such systems. Suitably induced spatio-temporal variations of active parameters combined with machine learning techniques could enable efficient control of complex flows and pattern formation in active microfluidic devices~\cite{wu2017transition,giomi2017cross,green2017geometry,zhang2019structuring,Peng2016command,vcopar2020microfluidic} and biological systems~\cite{saw2017topological,duclos2017topological}. Beyond active and soft matter, our deep neural-network models could be employed in other contexts where coupled chaotic fields naturally occur such as turbulent flows or magneto-hydrodynamics.~\cite{ling2016reynolds,duraisamy2019turbulence}

\vspace{10mm}

\section*{Methods}

\subsection*{Active Nematohydrodynamics and Lattice Boltzmann Simulation}

Simulation data for training and testing was generated using a hybrid lattice Boltzmann method which has been used in prior studies of active nematics \cite{Zhang2017,Kumar2018,zhang2019structuring}.
The symmetric and traceless tensorial order parameter of the nematic is defined as \begin{equation}
    {\bf Q}=S({\bf n}{\bf n}-{\bf I}/3)
    \label{eq:order_parameter}
\end{equation}
with $S$ being the scalar order parameter, ${\bf n}$ being the unit vector describing the local nematic orientation, and ${\bf I}$ being an identity tensor. The following governing equation of the nematic microstructure, namely Beris-Edwards equation (\ref{eq:berisedwards}) reads
\begin{equation}
    (\partial_t + \mathbf{u} \cdot \nabla) \mathbf{Q} - \mathbf{S} ( \mathbf{W}, \mathbf{Q}) = \Gamma \mathbf{H}
    \label{eq:berisedwards}
\end{equation}
where ${\bf u}$ is the velocity vector and $\Gamma$ is related to the rotational viscosity $\gamma_1$ via $\Gamma=2S^2/\gamma_1$ Here, the generalized advection term $\mathbf{S} ( \mathbf{W}, \mathbf{Q})$ is defined as

\begin{equation}
\begin{aligned}
\mathbf{S} ( \mathbf{W}, \mathbf{Q}) &= (\xi {\bf A} + {\bf \Omega})({\bf Q} + {\bf I}/3) \\
& + ({\bf Q} + {\bf I}/3)(\xi {\bf A} - {\bf \Omega}) \\
& - 2\xi ({\bf Q}+{\bf I}/3))\Tr({\bf QA})
\end{aligned}
\label{eq:advection}
\end{equation}

with ${\bf A}=(\nabla{\bf u}+(\nabla{\bf u})^T)/2$ being the strain rate tensor, ${\bf \Omega} = (\nabla{\bf u} - (\nabla{\bf u})^T)/2$ being the vorticity, and $\xi$ being a flow-alignment parameter setting the Leslie angle. The molecular field ${\bf H}$ is a symmetric, traceless projection of the functional derivative of the free energy of the nematic. Its index form reads

\begin{equation}
H_{ij}=\frac{1}{2}\left( \frac{\delta F}{\delta Q_{ij}} + \frac{\delta F}{\delta Q_{ji}} \right) - \frac{\delta_{ij}}{3} \Tr(\frac{\delta F}{\delta Q_{ij}})
\end{equation}
in which the free energy functional is $F=\int_V f dV$. Its density $f$ takes the following form:

\begin{equation}
\begin{aligned}
f&=\frac{A_0}{2}\left( 1-\frac{U}{3} \right) Q_{ij}Q_{ij} - \frac{A_0U}{3} Q_{ij}Q_{jk}Q_{ki} \\
& + \frac{A_0U}{4}Tr(Q_{ij}Q_{ij})^2+\frac{1}{2} L \partial_k Q_{ij} \partial_k Q_{ij}
\end{aligned}
\end{equation}

where $A_0$, $U$ are material constants and $L$ is related to the Frank elastic constant under the one-constant-approximation. Eq.\ref{eq:berisedwards} is solved using a finite difference method.

The hydrodynamic flow is governed by a momentum equation:
\begin{equation}
\begin{aligned}
   & \rho (\partial_t + u_j \partial_j) u_i = \partial_j \Pi_{ij} \\
   & + \eta \partial_j \left[ \partial_i u_j + \partial_j u_i +  (1 - 3 \partial_{\rho} P_0) \partial_{\gamma} u_{\gamma} \delta_{ij} \right]
    \label{eq:momentum}
\end{aligned}
\end{equation}
where $\rho$ is density, $\eta$ is the isotropic viscosity, and $P_0=\rho T - f$ is the hydrostatic pressure with $T$ being the temperature. The additional stress has two contributions, $\Pi_{ij}=\Pi_{ij}^p+\Pi_{ij}^a$, where the first term is passive in its nature accounting for the anisotropy, and is defined as
\begin{equation}
\begin{aligned}
\Pi_{ij}^p = & -P_0 \delta_{ij} - \xi H_{ik} \left( Q_{kj} + \frac{1}{3} \delta_{kj} \right) \\
& - \xi \left( Q_{ik} + \frac{1}{3} \delta_{ik} \right) H_{kj} \\
& + 2\xi \left( Q_{ij} + \frac{1}{3} \delta_{ij} \right) Q_{kl} H_{kl} \\
& - \partial_j Q_{kl} \frac{\delta F}{\delta \partial_i Q_{kl}} +Q_{ik} H{kj} - H_{ik} Q_{kj}
\end{aligned}
\end{equation}
The active stress that drives the system out-of-equilibrium reads
\begin{equation}
\Pi_{ij}^a = - \alpha Q_{ij}
\end{equation}
in which $\alpha>0$ describes an extensile active nematic, as is the case for the experimental systems discussed in this manuscript. Eq.\ref{eq:momentum} is solved simultaneously via a lattice Boltzmann method over a D3Q15 grid \cite{Guo2002}. Additional details on this method can be found in \cite{Zhang2016a}.

Typical simulation parameters were $\Gamma = 0.13$, $\eta = 0.33$, $A = 0.1$, and $U = 3.5$, leading to $q_0 \sim 0.62$. For Figs.~\ref{fig:parameter_estimation},~\ref{fig:time_evolution}, simulation were trained on $K=0.075$, $\alpha \in [0, 0.05]$. The range of $K$ for testing in Fig.~\ref{fig:parameter_estimation}f,g was $K \in [0.06, 0.09]$. For the multiparameter estimator used in Figs.~\ref{fig:experimental_data},~\ref{fig:time_dependence},Supplementary Fig.~\ref{fig:actin-multi2d}, $K \in [0.06, 0.20]$ and $\alpha \in [0, 0.05]$. For the experimental prediction in 3D (Supplementary Fig.~\ref{fig:3d_predictions}), models were trained on $K=0.1$, $\alpha \in [0, 0.09]$ as initial predictions indicated that the range of $\alpha$ first used for training was insufficient.

\subsection*{Machine Learning Details}

Neural networks are implemented in Python using the Pytorch library. Code for data preparation, network implementation, training, and evaluation is available at \url{https://github.com/jcolen/active_nematics}

\subsubsection*{\textbf{Parameter estimation}}

Parameter estimation networks contain between 1-2 convolutional layers with hyperbolic tangent activation functions, each of which is followed by a max pooling layer and a dropout layer with dropout probability of 0.15. 
The convolutional layers are further connected with a single recurrent layer implemented with a long short-term memory cell. 
Lastly, a dense layer with linear activation function is added to output the predicted parameters.
An example architecture is shown in Fig.~\ref{fig:parameter_estimation}e. 
To make predictions on large director field images, the network randomly selects patches and ensemble averages the results into a final prediction.
For networks using a recurrent layer, the model accepts a sequence of director field frames, rather than a single frame. 

Three parameter estimation models are used in this paper. The first, used to predict activity in two-dimensional nematics (Fig.~\ref{fig:parameter_estimation}f and Fig.~\ref{fig:time_dependence}), has a single convolutional layer with 32 filters of size $3 \times 3$, a single $2\times2$ max pooling layer, a recurrent layer implemented using a long short-term memory (LSTM) with hidden size 32, and a fully-connected layer with 32 neurons. This model accepts input sequences of $32\times32$ pixel image patches and was trained on a dataset of 6,000 director field frames separated by 10 simulation time steps, at 12 different levels of activity. The second model, used to predict activity in three-dimensional nematics (Fig.~\ref{fig:parameter_estimation}g, Supplementary Fig.~\ref{fig:3d_predictions}) has a similar architecture, but with $5\times5\times5$ convolutional filters and no recurrent layer. This model accepts image volumes of size $32\times32\times32$ and was trained on a dataset of 6,000 director field configurations, separated by 100 time steps, at 12 levels of activity. The third model is used for simultaneous prediction of activity and elastic modulus in two-dimensional nematics (Fig.~\ref{fig:experimental_data} and Supplementary Fig.~\ref{fig:actin-multi2d}). This network has the same structure as the other two-dimensional model, but outputs two values and was trained on a dataset of 15,000 image frames, generated with 30 different combinations of activity and elastic modulus. The accuracy of this multi-parameter estimator is summarized in Fig.~\ref{fig:multi2d}. 

Networks were trained for 100 epochs on director field configurations generated using Lattice-Boltzmann simulations. Each frame of training data was a $200\times200$ director field image with periodic boundary conditions. These datasets were augmented by applying random rotations, flips, and shifts during the training procedure. During each epoch, each input frame was randomly cropped to the predictive network input size. During training, we used an 80-20 training-validation split on the input dataset.

\subsubsection*{\textbf{Predicting Time Evolution}}

\paragraph*{\textbf{Autoencoder architecture}}

The neural network for predicting time evolution, depicted in Fig.~\ref{fig:time_evolution}a, is comprised of three parts: an encoder, recurrent layers, and a decoder. The encoder uses a sequence of convolutional layers to downsample input images into feature vectors. The decoder accepts feature vectors and uses convolutional layers to upsample those feature vectors back into images. A traditional autoencoder is comprised of these two layers only, and is an effective method of reducing data dimensionality. In our model, we insert the recurrent layers in between the encoder and decoder, so that dynamics can be computed on the encoded feature vectors. A benefit of this approach is that the dimensional reduction achieved by the encoder allows for smaller recurrent layers, reducing network complexity and improving performance. 

The models reported in this paper accept director field images processed into the 2-channel input $(\sin(2\theta), \cos(2\theta))$, where $\theta$ is the local orientation angle of the director field. The encoder contains two convolutional layers of stride-2 with 4 and 6 output channels, respectively. The decoder architecture mirrors that of the encoder, accepting a 6-channel feature vector and using two stride-2 transposed convolutional layers with 4 and 2 output channels, respectively. All convolutional layers use $4\times4$ kernels and are followed by batch normalization, which improves training performance, as well as hyperbolic tangent activation. The recurrent portion is a two-layer long short-term memory (LSTM) unit implemented as a residual network, or resnet, with a shortcut that directly connects input and output of the entire LSTM cell. Given a sequence of feature vectors, the resnet computes a small residual to be added to the input, rather than computing a full output feature vector from scratch. For input sequences with small time separations, the residual vector is sparse, which helps improve training performance and predictive accuracy. 

These models were trained using a two-step training procedure. First, the encoder, resnet, and decoder were trained together for 100 epochs. Next, the weights in the resnet were frozen and the encoder and decoder were trained together for 50 epochs. Training data was generated either using Lattice-Boltzmann simulations or directly from experiments. The Lattice-Boltzmann training data consisted of $200\times200$ director field images with periodic boundary conditions, separated by 6, 10, and 25 simulation time steps. Each simulation dataset contained 6,000 director field configurations at 12 levels of activity and was augmented during training using random flips, shifts, and crops. As before, we used an 80-20 training/validation split. Different models were trained on each dataset, with input image sizes of $48\times48$, $64\times64$, and $120\times120$. In the main text, we report results from the best performing of these models, which were trained on data with a frame separation of 10 time steps and use $48\times48$ input image size. 

The experimental data consisted of 1,500 director field configurations extracted from microtubule-kinesin experiments at 5 different ATP concentrations (see Experimental Methods). We did not train on experiments with ATP concentrations of 10 $\mu$M and 18 $\mu$M as the time between snapshots was $5\times$ longer than for the other ATP concentrations. Here, we also used an 80-20 training validation split and augmented data using random flips and crops. The results reported in this paper are for a model with an input size of $48\times48$.
\vspace{5mm}

\paragraph*{\textbf{Stitching predictions}}

While the models were trained to predict the evolution of director field patches, the error rates and characteristic length and time scales reported in Fig~\ref{fig:time_evolution} are computed for full images in the testing dataset. To obtain the predicted configuration of the full director field, the model stitches together predictions made on overlapping subdomains of the image. Here, each pixel will appear in the prediction for multiple subdomains. The final prediction for each pixel is given by the weighted average of predictions from each subdomain. For a pixel located at $\mathbf{r} = (x, y)$, the weight given to the predicted value from a subdomain centered at $\mathbf{r_0} = (x_0, y_0)$ is the Gaussian weight with $\sigma = R$, the radius of the subdomain. Thus, more credence is given to domains in which the pixel is farther from the boundary. For all results reported in this paper, predictions were stitched together from 48$\times$48 ($R = 24\sqrt{2}$) domains which overlapped by 8 pixels.
\vspace{5mm}

\paragraph*{\textbf{Sharpening algorithm}}
The sharpening procedure is written in Python using the Numba library and executes elastic free energy minimization. 
Following \cite{deGennes1995Crystals,Chaikin1995PrinciplesPhysics}, we write the elastic free energy density as

\begin{equation}
\begin{aligned}
    f_d &= \frac{1}{2} K_1 (\nabla \cdot \mathbf{n})^2 + \frac{1}{2} K_2 (\mathbf{n} \cdot \nabla \times \mathbf{n})^2 \\
    &+ \frac{1}{2} K_3 (\mathbf{n} \times (\nabla \times \mathbf{n}))^2 
\end{aligned}
\end{equation}

Assuming a two dimensional system parameterized by an angle $\theta$ as $\mathbf{n} = (\cos \theta, \sin \theta)$, this becomes:

\begin{equation}
\begin{aligned}
    f_d &= \frac{1}{2} K_1 (\sin \theta \partial_x \theta - \cos \theta \partial_y \theta)^2 \\
    &+ \frac{1}{2} K_3 (\cos \theta \partial_x \theta + \sin \theta \partial_y \theta)^2
\end{aligned}
\end{equation}

In the one elastic constant approximation $K_1 = K_3 = K$, this reduces to

\begin{equation}
    f_d = \frac{1}{2} K \left[(\partial_x \theta)^2 + (\partial_y \theta)^2 \right]
\end{equation}

The elastic free energy is minimized by setting $\frac{\delta f_d}{\delta \theta} = 0$, leading to the Laplace equation:

\begin{equation}
    \nabla^2 \theta = 0
\end{equation}

Thus, the elastic free energy minimization can be accomplished by applying relaxational dynamics to the director field. We implement this using a standard finite-differences approach, slightly modified to account for the nematic symmetry $\mathbf{n} = -\mathbf{n}$.

We first apply relaxational dynamics in a small box surrounding the topological defect positions from the previous director field frame. Because the winding number around the boundary of this box is fixed and nonzero, this sharpens the director field around each defect without risking removing the defect. Next, the director field is fixed inside the box and relaxational dynamics are applied in the defect-free region. This procedure is applied iteratively to sharpen the raw predicted image. 

This procedure will work if the defect has not moved outside of the box between image frames. Assuming a timestep of $\tau$, box size of $R$, and characteristic defect velocity $v_d$, this condition is satisfied if $v_d \tau < R$. We can approximate $v_d$ using the relation provided by \cite{Giomi2014} for an isolated $+1/2$ defect, $v_d \approx \alpha l_d / \eta$. Here, we insert $l_d \approx \sqrt{K / \alpha}$, the mean defect spacing, as the radius of the defect-free region surrounding the $+1/2$ defect. Thus, the defect will remain in the box if $\sqrt{\alpha K} \tau / \eta < R$. The simulation data used in Fig.~\ref{fig:time_evolution} had $K = 0.075$, $\eta = 0.33$, $\alpha_{\text{max}} = 0.05$, and $\tau = 10$, leading to $R > 2$. The data reported in the main text was generated using a 5$\times$5 box, corresponding to $R \in [2.5, 3.5]$. We chose the smallest possible value for $R$ above this threshold, as it prevented the immediate annihilation of recently-nucleated defect pairs, which would otherwise be close enough to be enclosed by the same box. This would result in a net zero winding around the boundary, leading to their removal by the sharpening procedure.

\subsubsection*{\textbf{Applications to experiment}}

Experimental data is pre-processed before being fed through parameter prediction models. The data is first adjusted in ImageJ to remove outliers using a median filter, and then downsampled. For actin-myosin, we downsample by a factor of 6 to an effective pixel width of $a=1$ $\mu$m, a convention that has been used in the past when comparing this Lattice-Boltzmann code with actin-myosin nematics \cite{Zhang2017,Kumar2018}. The microtubule-kinesin data was downsampled by a factor of 16 to an effective pixel-width of $a=5.2$ $\mu$m, as the length scale of spatial variations in the raw data is larger.

\subsection*{Determination of Lyapunov Time}

Active nematics are a nonlinear system characterized by a positive Lyapunov exponent. As a result, direct comparison of time-evolved director field configurations is not necessarily valid for long timescales. Pixel-wise accuracy should not be expected beyond the Lyapunov time, particularly as our predictive model lacked complete information about the system. While we report pixel-wise accuracy in the main text (Fig~\ref{fig:time_evolution}b), knowledge of the chaotic dynamics of these systems is important to contextualize these results. 

The grayed out region in Fig.~\ref{fig:time_evolution}b is bounded by the Lyapunov time as determined from Lattice-Boltzmann simulations. To find this timescale, we ran Lattice-Boltzmann simulations at different levels of activity and saved an intermediate state of the system. We then perturbed this state and continued the simulation. At each level of activity, we ran 10 trials from 10 separate intermediate states. Each simulation time-evolved at 200x200 grid, from which 100 points were randomly selected and tracked over time. By comparing these randomly selected pixels as a function of time, we extracted the Lyapunov exponent which was inverted to obtain the Lyapunov time. This quantity was dependent on activity, with more active systems exhibiting a shorter Lyapunov time. However, when we rescaled by the characteristic defect lifetime $\tau_d = \eta / \alpha$, we found that all values coalesced to approximately $\tau = 3.6 \tau_{\text{d}}$. 

As infinitesimal pixel-wise changes would be eliminated by relaxational dynamics, we used a more global method of perturbing the intermediate state. First, we computed the singular value decomposition of the order parameter tensor field $Q_{ij} (\mathbf{r})$. We then fractionally changed 10 elements of the singular matrix by random amounts between -10\% and +10\% and used the new matrix to reconstruct the perturbed order parameter field. This method of globally varying the intermediate state yielded non-vanishing pixel-wise deviations that showed exponentially growing behavior. 

\subsection*{Characteristic Length and Time Scales}

Direct comparisons of the machine learning predicted director field to Lattice-Boltzmann simulations are unreliable beyond the Lyapunov time. To evaluate the validity of our predictions over longer time scales,  we compare instead characteristic length and time scales of the machine learning predicted dynamical steady state. For a given order parameter configuration $Q_{ij}(\textbf{r}, t)$, with $i$ and $j$ running over $x, y$, we define the spatial correlation function $C_s(\textbf{r}, t)$ as
\begin{align}
    C_s(\mathbf{r}, t) = \frac{\left< Q_{ij}(\mathbf{r}, t) Q_{ij}(0, t) \right>}{\left< Q_{ij}(0, t) Q_{ij}(0, t) \right>} 
    \label{eq:corr_l}
\end{align}
and the time correlation function $C_t(\mathbf{r}, t)$
\begin{align}
    C_t(\mathbf{r}, t) = \frac{\left<Q_{ij}(\mathbf{r}, t) Q_{ij}(\mathbf{r}, 0)\right>}{ \left<Q_{ij}(\mathbf{r}, 0) Q_{ij}(\mathbf{r}, 0)\right>}
    \label{eq:corr_t}
\end{align}
where indices $i$ and $j$ are contracted following the Einstein summation convention. Using Eqs. (\ref{eq:corr_l}) and (\ref{eq:corr_t}), we define the director field correlation length $\ell_{\theta}$ such that $C_s(\ell_{\theta}, t) = 1/2$ and the correlation time $t_{\theta}$ such that $C_t(r, t_{\theta}) = 1/2$.

In Fig.~\ref{fig:time_evolution}c,d, we compare the average values of $\ell_{\theta}$, $t_{\theta}$ as found in machine learning predicted director field frames to those of Lattice-Boltzmann simulations. Here, we iterate the predictive model to predict large ($200\times200$) image frames over a long time ($t = 50\ \tau_{\text{LC}}$) and compute the time-averaged correlation length and spatially-averaged correlation time. In Supplementary Fig.~\ref{fig:mean_defect_spacing}, we report the time-averaged mean-defect spacing, defined as $\ell_{\text{d}} = 1 / \sqrt{n_{\text{d}}}$, where $n_{\text{d}}$ is the defect density.  

\vspace{5mm}

\subsection*{Experimental Methods}

\subsubsection*{\textbf{Actin-Myosin Nematics}}

\begin{figure}[H]
    \centering
    \includegraphics[width=0.32\textwidth]{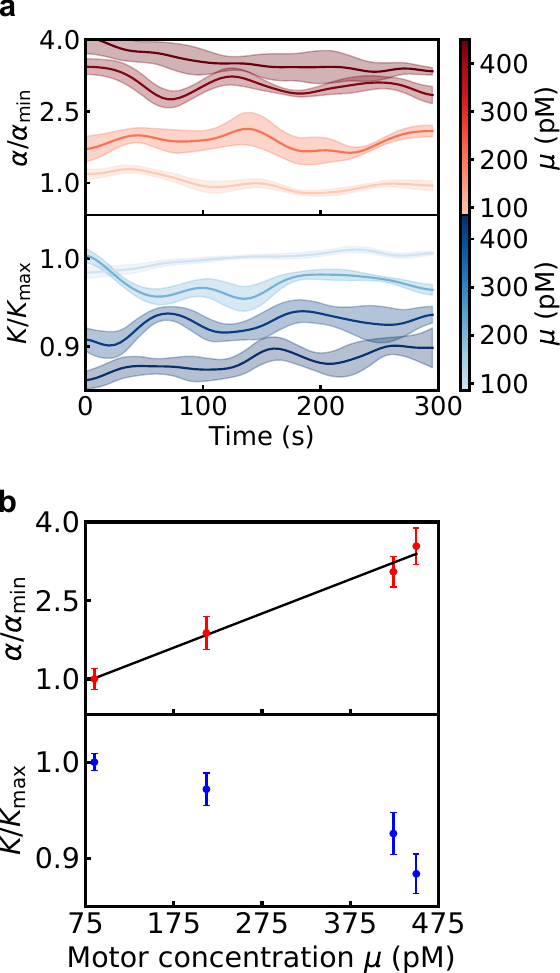}
    \caption{\textbf{Multiparameter estimation of actin-myosin.} \textbf{a.} Simultaneous prediction of $\alpha$, $K$ as a function of time and myosin motor concentration. Here $\alpha_{\text{min}}$, $K_{\text{max}}$ are the time-averaged predicted activity and elastic modulus at the lowest level of motor concentration $\mu_{\text{min}} = 85$ pM. \textbf{b.} Time-averaged machine learning predictions of $\alpha$, $K$ as a function of motor concentration. The fit uses a linear scaling proposed by Ref.~\cite{Kumar2018}.}
    \label{fig:actin-multi2d}
\end{figure}

Experiments for Fig.~\ref{fig:time_dependence}d and Supplementary Fig.~\ref{fig:actin-multi2d} were performed as in ~\cite{zhang2019structuring} using the method originally described in ~\cite{Kumar2018}; the experimental data for Fig. ~\ref{fig:time_dependence}e-f is taken directly from ~\cite{zhang2019structuring}.  Supplementary Table \ref{tab:actin} contains a full enumeration of the conditions in each experiment but the general method will be summarized here. Fluorescent (TMR labelled) and non-fluorescent actin monomers are mixed at a ratio of 1:5 and allowed to polymerize in F-buffer (1 mM MgCl2, 50 mM KCl, 0.2 mM egtazic acid (EGTA)) containing either imidazole (10mM, pH 7.5) or HEPES (10mM, pH 7.5) as a buffering reagent. Also present is f-actin capping protein to limit the length of nascent filaments to ~2 $\mu$m. Additionally the mix contains a Glucose Oxidase/Catalase oxygen scavenging system (2.7 mg/ml glucose oxidase, 1700 U/ml catalase, 4.5 mg/ml glucose, 0.5\% v/v $\beta$-mercaptoethanol) to limit photo damage. Lastly, the polymerization mix contains methylcellulose (0.3\% weight/volume in water) as a crowding agent and ATP as a source of chemical fuel. 

The experiment is performed in a chamber composed of a glass cloning cylinder attached to a glass coverslip via 5 minute epoxy. Before adding the sample to the chamber, the bottom is first coated with a thin layer of oil that contains a surfactant (PFPE-PEG-PFPE surfactant) to prevent the filaments from sticking to the glass. The sample is allowed to settle as the methylcellulose crowds the filaments to the oil-water interface, forming a nematic liquid crystal. The sample is imaged on a inverted spinning disk confocal microscope. 

Activity is introduced via the addition of synthetic myosin motors after the sample has formed a liquid crystal at the oil-water interface. Two different motor constructs were used. The synthetic motors used for Supplementary Fig.~\ref{fig:actin-multi2d} are the engineered myosin tetramers CM11CD$_{746}$2R{\raise.17ex\hbox{$\scriptstyle\sim$}}1R{\raise.17ex\hbox{$\scriptstyle\sim$}}TET from Ref ~\cite{Schindler2014}, referred to in this study as "fixed-gear motor". These motors are constructed from the catalytic domain of a fast algal myosin (\textit{Chara corallina} myosin XI) fused to an artificial lever arm consisting of 2 spectrin repeats. The motors oligomerize into tetramers with flexible linkages using the engineered leucine zipper variant pLI2 and contain a spectrin repeat flanked by flexible linkers. The synthetic motor used for the experiment in  Fig.~\ref{fig:time_dependence}e-f is MyLOVChar4{\raise.17ex\hbox{$\scriptstyle\sim$}}1R{\raise.17ex\hbox{$\scriptstyle\sim$}}TET \cite{Ruijgrok2020}, referred to in this study as "gearshifting" motor, which is the same construct used for experiments in \cite{zhang2019structuring} including the data from that study analyzed in Fig. ~\ref{fig:time_dependence}e-f.   This motor construct is also based on the catalytic domain of \textit{Chara corallina} myosin XI, but utilizes an artificial lever arm that contains the light activatable LOV2 domain that unfolds upon stimulation with blue light. The conformational change of the lever arm results in a light-dependent stroke vector for the motor \cite{Ruijgrok2020} which,in the context of the active nematic system in this study, results in higher activity in the presence of blue light \cite{zhang2019structuring}. The fixed-gear and gearshifting motors were purified, flash frozen and stored at -80 Celsius as described \cite{Ruijgrok2020}. Time varying activity movies were stimulated over the entire sample via a 491 nm solid state laser, while spatially varying activation was achieved by targeting a 470 nm LED to one location in the sample using a digital micromirror array ~\cite{zhang2019structuring}.

\subsubsection*{\textbf{Microtubule-Kinesin Nematics}}

The microtubule and kinesin motor based active nematics were assembled according to previously published methods \cite{sanchez2012spontaneous, decamp2015orientational}. Briefly, K401-BIO-HIS purified from \textit{E. Coli} \cite{Subramanian2007} was incubated with streptavadin for 30 min to create motor clusters. A mixture containing salt (5 mM MgCl$_2$), an ATP regeneration system (26.6 mM phosphenol pyruvate (Beantown Chemicals), pyruvate kinase/lactic dehydrogenase), the depletion agent (0.8\% w/v 20 kDa polyethylene glycol) and an anti-oxidant system (6.7 mg/mL glucose, 0.4 mg/mL glucose catalase, 0.08 mg/mL glucose oxidase and 2mM trolox) was combined with the kinesin motor clusters in a buffer of M2B (80 mM PIPES, pH 6.8, 1 mM EGTA, 2 mM MgCl$_2$). The ATP was added in the desired concentration (from 5 $\mu$M to 1000 $\mu$M) to individual aliquots before flash freezing in liquid nitrogen. The concentration of ATP controls the stepping rate of the motors, which in turn tunes the level of activity in the system. 

The tubulin was purified from bovine brain \cite{PopovTubulin}. Tubulin labeled with NHS-Alexa 647 \cite{HymanTubulin} was co-polymerized with unlabeled tubulin in the presence of Guanosine-5-[($\alpha,\beta$)-methyleno]triphosphate GMPCPP (Jena Bioscience NU-405L) to create microtubules with a final fraction of 3\% labeled tubulin and an average length of 2.5 $\mu$m. The activity in the nematics is sensitive to the particular protein preparation, MT length distribution and the chemical properties of the active mixture. Therefore it was important to use stocks from the same preparation and polymerization to get quantitatively consistent results. 

The experiments were performed in a flow chamber with dimensions 18 $\times$ 3 $\times$ 0.6 mm made of double-sided tape sandwiched between a glass slide and a coverslip. The glass slide was treated with commercially available Aquapel to create a hydrophobic surface. The coverslip was passivated with acrylamide. 

On the day of experiments, the pre-mixed active components and MTs were thawed rapidly and combined. To form a large flat interface, first oil (HFE 7500) stabilized with a fluoro-surfactant PFPE-PEG-PFPE (1.8\% RAN Biotech) was flowed into the chamber and then the aqueous active mixture. The hydrophobic treatment left a thin layer of oil for the MTs to sediment onto. The chamber was sealed with Norland Optical Adhesive and cured under a UV light for 1 min. The sedimentation of the microtubules onto the oil-water interface was aided by spinning the sample in a swinging bucket centrifuge (Sorval Legend RT rotor \#6434) at 1000RPM for 20 min. 

The samples were imaged using epi-fluorescence microscopy on a Nikon Ti-Eclipse equipped with a CMOS camera (Andor Zyla). The orientation field was extracted from the fluorescence images using the Image-J plugin Orientation-J which finds that structure tensor from gradients in intensity.

\onecolumngrid
\clearpage
\newpage
\twocolumngrid

\section*{Supplementary Figures}

\begin{figure}[H]
    \centering
    \includegraphics[width=0.49\textwidth]{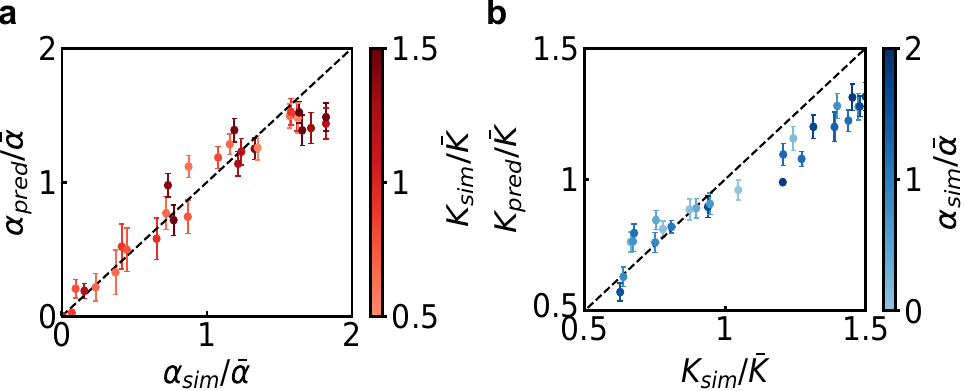}
    \caption{\textbf{Multiparameter Estimation in Simulations.} Predictive accuracy of the multi-parameter estimator for predicting $\alpha$ (\textbf{a}) and $K$ (\textbf{b}) in 2D active nematics. Here, $\bar{\alpha}$ and $\bar{K}$ are the mean values of $\alpha$ and $K$ from the training set. Results of this model applied to experimental systems are shown in Fig.~\ref{fig:experimental_data} and Supplementary Fig.~\ref{fig:actin-multi2d}.}
    \label{fig:multi2d}
\end{figure}

\begin{figure}[b]
    \centering
    \includegraphics[width=0.32\textwidth]{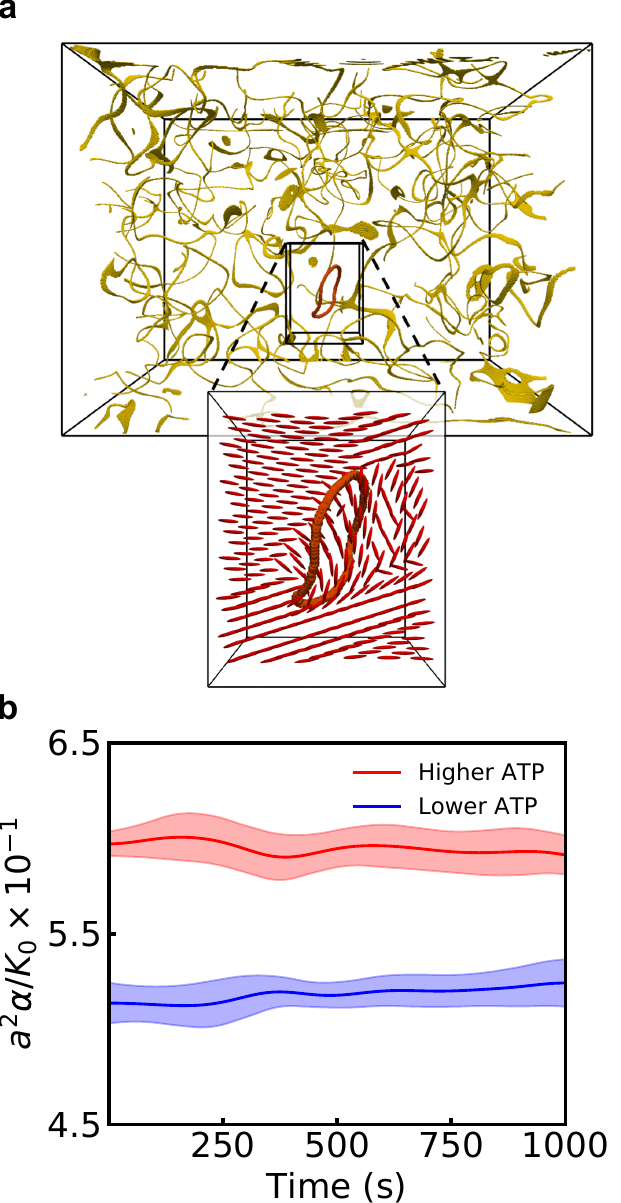}
    \caption{\textbf{3D microtubule-kinesin active nematics experiments.} \textbf{a.} Disclination loop structure of an experimental realization of a three-dimensional active nematic. Inset shows a zoom of the director field configuration near a disclination loop. \textbf{b.} Machine learning predicted activity over time for two experimental configurations at different ATP concentrations.}
    \label{fig:3d_predictions}
\end{figure}

\begin{figure}[b]
    \centering
    \includegraphics[width=0.36\textwidth]{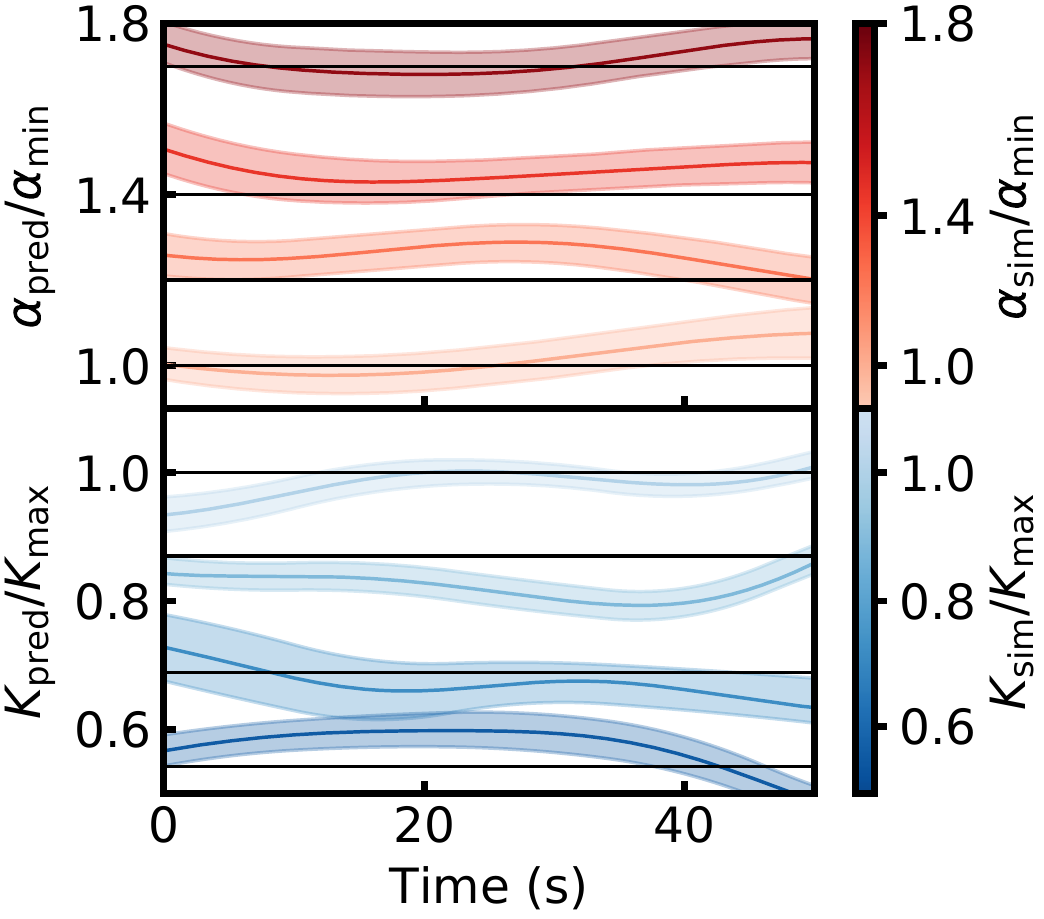}
    \caption{\textbf{Intrinsic temporal fluctuations of the machine learning based parameter estimation.} Machine learning predictions of $\alpha$ and $K$ for Lattice-Boltzmann simulations performed at constant parameter values are shown as a function of time. Owing to the machine-learning uncertainty, the predicted values (colored curves) fluctuate near the reference levels (black lines) that are prescribed in simulation.}
    \label{fig:fluctuations_const}
\end{figure}

\begin{figure}[H]
    \centering
    \includegraphics[width=0.32\textwidth]{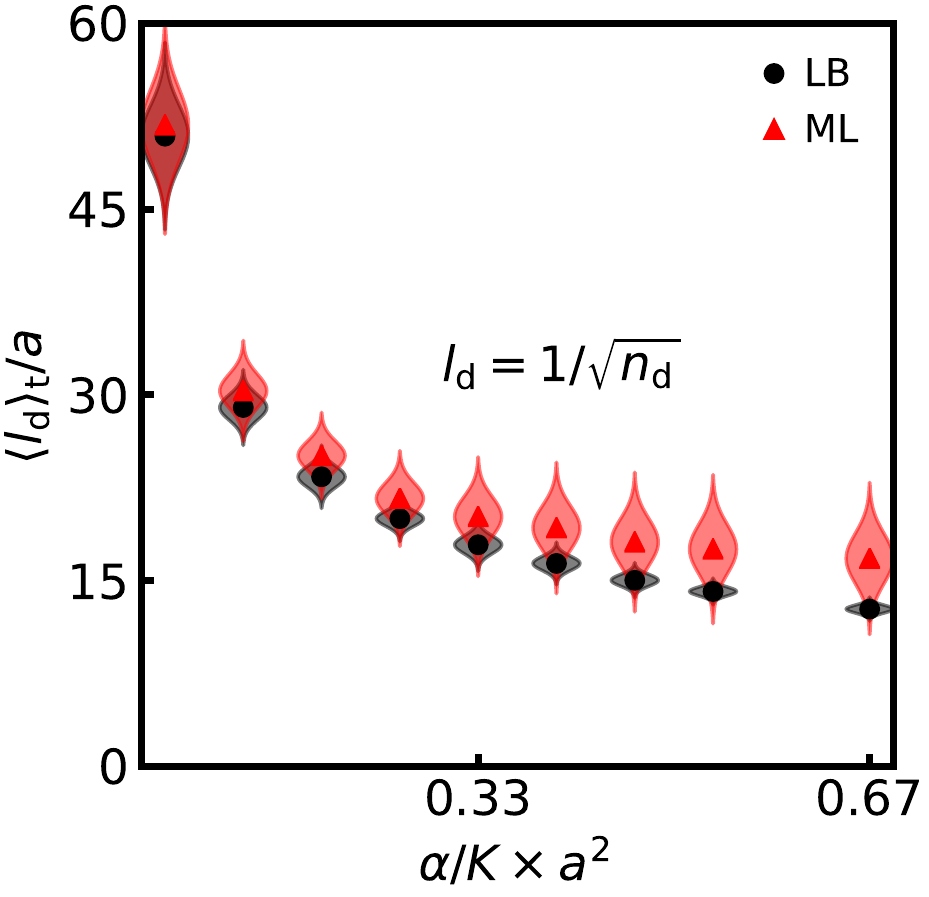}
    \caption{\textbf{Mean defect spacing.} Comparison of time-averaged mean defect spacing in machine learning and Lattice-Boltzmann simulations.}
    \label{fig:mean_defect_spacing}
\end{figure}

\begin{figure}[H]
    \centering
    \includegraphics[width=0.3\textwidth]{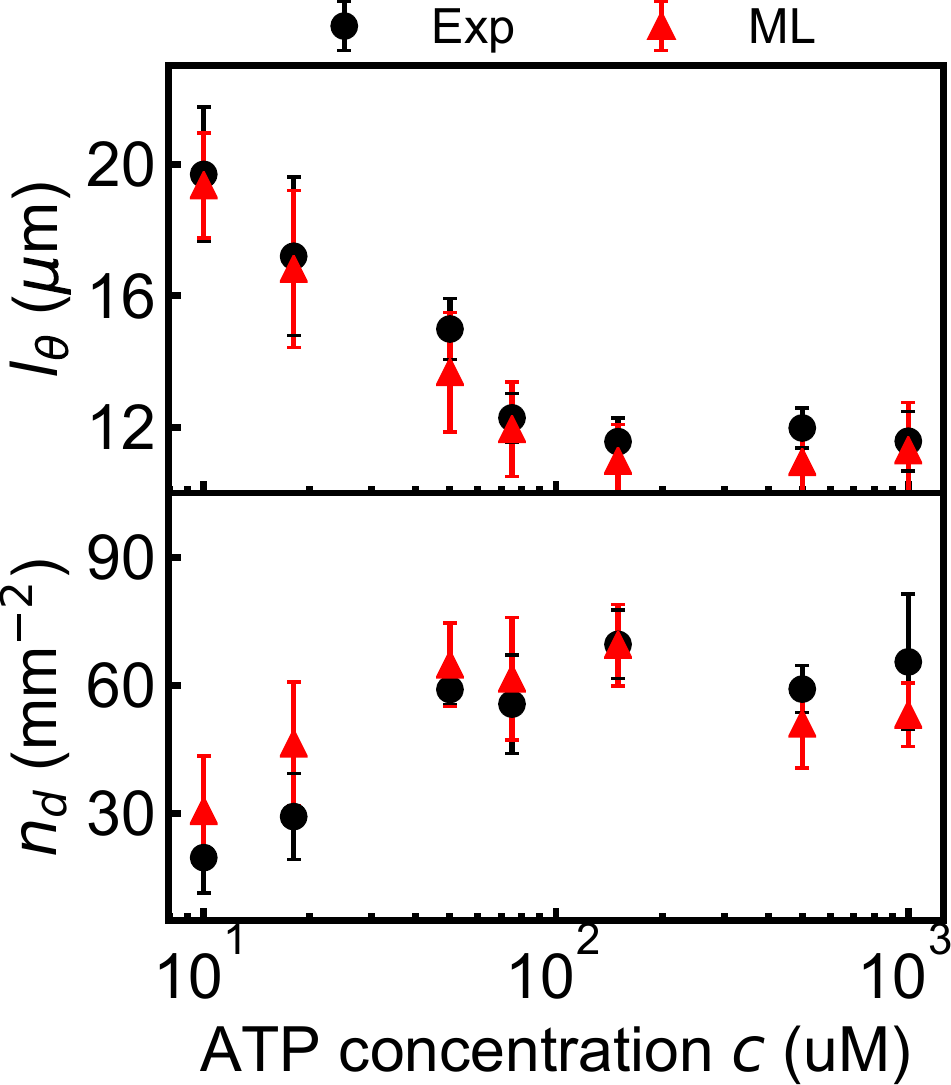}
    \caption{\textbf{Machine learning predicted length scales in microtubule-kinesin nematics.} Comparison of director field field correlation length $\ell_{\theta}$ and mean defect spacting $n_{\text{d}}$ in experiments (Exp) and as predicted by our machine learning time evolution algorithm (ML) over a range of ATP concentrations.}
    \label{fig:mt_length_scales}
\end{figure}

\begin{enumerate}
    \item Figure \ref{fig:multi2d} - Multiparameter machine learning estimation accuracy for two-dimensional active nematics. This is the multi-parameter model used to predict $\alpha, K$ in Fig. \ref{fig:experimental_data}.
    \item Figure \ref{fig:3d_predictions} - Predicted activity in 3D microtubule-kinesin active nematics.
    \item Figure \ref{fig:actin-multi2d} - Simultaneous estimation of $\alpha, K$ in actin-myosin systems at various motor concentrations. 
    \item Figure \ref{fig:fluctuations_const} - Machine learning predicted activity over time in Lattice-Boltzmann simulations with constant activity
    \item Figure \ref{fig:defect_nucleation} - Comparison of a predicted defect nucleation event for Lattice-Boltzmann simulations and machine learning predictions.
    \item Figure \ref{fig:mean_defect_spacing} - Comparison of time-averaged mean defect spacing in machine learning and Lattice-Boltzmann simulations.
    \item Figure \ref{fig:mt_length_scales} - Comparison of characteristic length scales in microtubule-kinesin experiments and machine learning predictions of time evolution using those experiments across a range of ATP concentrations.
\end{enumerate}

\vspace{10mm}

\section*{Supplementary Movies}

\begin{figure*}[t]
  \centering
    \includegraphics[width=0.8\textwidth]{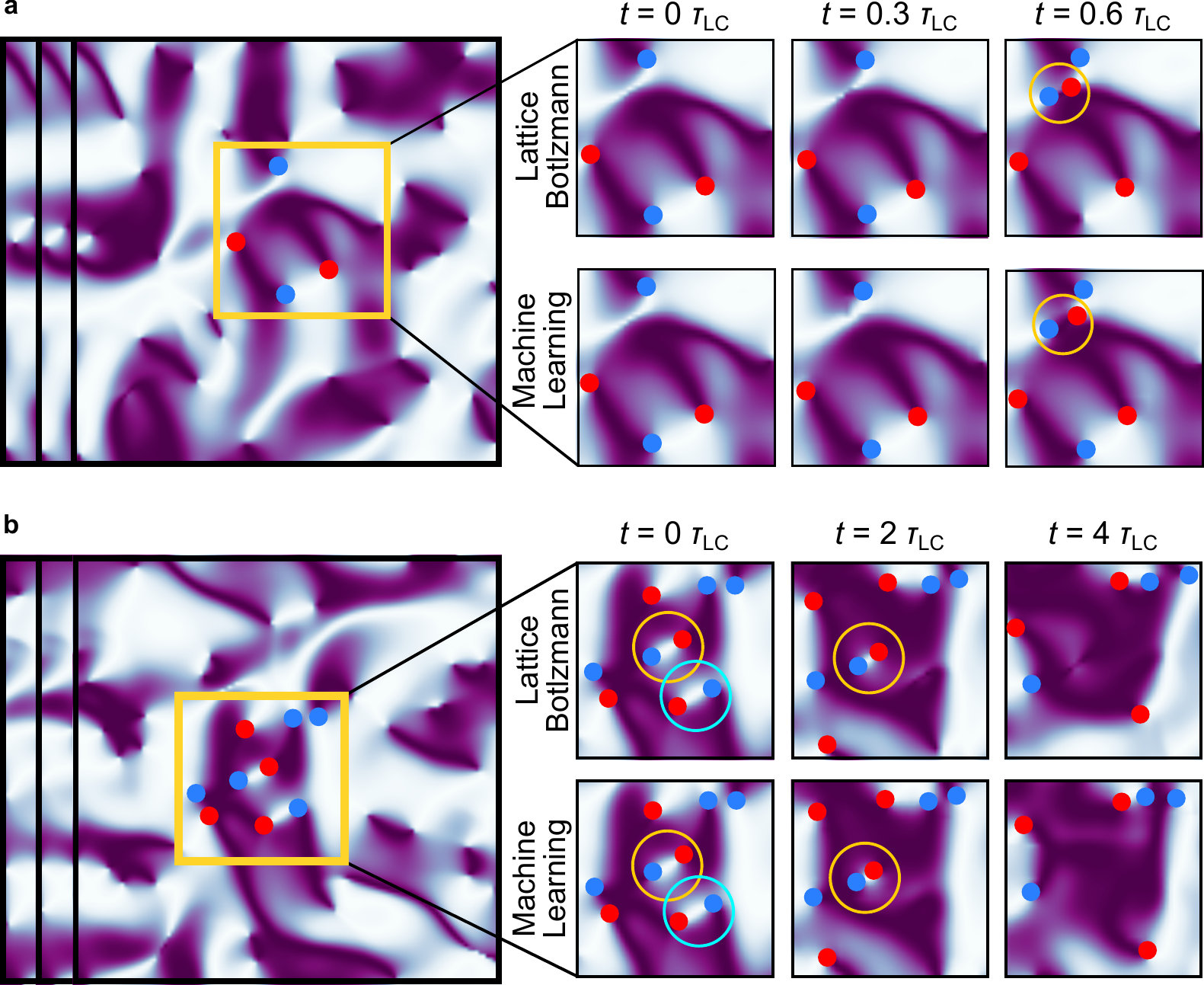}
    \caption{\textbf{Defect events.} \textbf{a, b.} Machine learning and Lattice Boltzmann predictions of defect nucleations (\textbf{a}) and annihilations (\textbf{b})}.
    \label{fig:defect_nucleation}
\end{figure*}

\begin{enumerate}
    \item Supplementary Movie 1 - Side-by-side comparisons of F-actin with light-activated myosin motors, and the spatio-temporal activity as predicted by machine learning. Part 1 depicts the data shown in Fig.~\ref{fig:time_dependence}d, in which the full field of view is illuminated at the indicated time. Part 2 depicts the data shown in Fig.~\ref{fig:time_dependence}e-f (data are from~\cite{zhang2019structuring}), with the selectively illuminated region denoted by dashed lines. 
    \item Supplementary Movie 2 - Side-by-side comparisons of defect nucleations and annihilations as predicted in lattice Boltzmann simulations and by our machine learning time-evolution model. Parts 1-2 depict the data shown in Fig.~\ref{fig:defect_nucleation}a-b, respectively. Data were generated at $\alpha = 0.02, 0.025, 0.03$, $K = 0.075$, with a frame separation of approximately 0.3 $\tau_{\text{LC}}$.
    \item Supplementary Movie 3 - Side-by-side comparisons of defect nucleations and annihilations as observed in microtubule-kinesin nematics experiments and as predicted by our machine learning time-evolution model. Part 1 depicts the data shown in Fig.~\ref{fig:time_evolution}e, with the machine learning model trained \textit{exclusively} on experimental data. Parts 2-4 show predictions on experiments by a model trained using simulation data. 
\end{enumerate}

\onecolumngrid
\vspace{10mm}
\section*{Supplementary Table}

\begin{table*}[h]
\label{t1}
\begin{center}
\begin{tabular}{ |p{1.8cm}|p{1.7cm}|p{1.5cm}|p{1.4cm}|p{1.5cm}|p{2cm}|p{1.7cm}| }
\hline
\textbf{Assay Conditions} & \textbf{F-buffer Imidazole} & \textbf{F-buffer HEPES} & \textbf{ATP} & \textbf{Capping Protein} & \textbf{Motor Concentration} & \textbf{Motor Type} \\

 \hline
Fig. ~\ref{fig:time_dependence}d &  & $\times$ & 1~mM & 20~nM & 200~pM & gearshifting \\ 
 \hline
Fig. ~\ref{fig:time_dependence}e  & $\times$ & & 100~$\mu$M & 16~nM  & 28 pM & gearshifting \\ 
 \hline
Fig. ~\ref{fig:actin-multi2d} & & $\times$ & 100~$\mu$M & 20~nM & 85, 212, and 424 pM & fixed gear \\ 
 \hline
Fig. ~\ref{fig:actin-multi2d}* &  & $\times$ & 1~mM & 20~nM & 450 pM & fixed gear \\ 
 \hline
\end{tabular}
\caption {Assay conditions used in Ref.~\cite{zhang2019structuring} (Fig.~\ref{fig:time_dependence}e) and in this study (all others) for experiments shown in the main and supplementary figures.}
\label{tab:actin}
\end{center}
\end{table*}

\clearpage
\twocolumngrid

\clearpage
\onecolumngrid

\twocolumngrid

\onecolumngrid
\clearpage
\newpage
\twocolumngrid

\bibliographystyle{naturemag.bst}
\bibliography{main}

\end{document}